\documentclass{article}

\usepackage{arxiv}

\usepackage[utf8]{inputenc} 
\usepackage[T1]{fontenc}    
\usepackage{hyperref}       
\usepackage{url}            
\usepackage{booktabs}       
\usepackage{amsfonts}       
\usepackage{nicefrac}       
\usepackage{microtype}      
\usepackage{lipsum}		
\usepackage{natbib}
\usepackage{amsmath}
\usepackage{amssymb}
\usepackage{subcaption}
\usepackage{bbm}
\usepackage{mathtools}

\title{Shrinkage Estimators in Online Experiments}

\author{
    Drew Dimmery\\
    Facebook\\
    Menlo Park, CA\\
    \texttt{drewd@fb.com}
    \And 
    Eytan Bakshy\\
    Facebook\\
    Menlo Park, CA\\
    \texttt{ebakshy@fb.com}
    \And 
    Jasjeet Sekhon\\
    UC-Berkeley\\
    Berkeley, CA\\
    \texttt{sekhon@berkeley.edu}
}

\DeclareMathOperator{\E}{\mathbb{E}}

\begin{document}
\maketitle

\begin{abstract}
We develop and analyze empirical Bayes Stein-type estimators for use in the estimation of causal effects in large-scale online experiments. 
While online experiments are generally thought to be distinguished by their large sample size, we focus on the multiplicity of treatment groups. 
The typical analysis practice is to use simple differences-in-means (perhaps with covariate adjustment) as if all treatment arms were independent. 
In this work we develop consistent, small bias,  shrinkage estimators for this setting. 
In addition to achieving lower mean squared error these estimators retain important frequentist properties such as coverage under most reasonable scenarios. 
Modern sequential methods of experimentation and optimization such as multi-armed bandit optimization (where treatment allocations adapt over time to prior responses) benefit from the use of our shrinkage estimators.
Exploration under empirical Bayes focuses more efficiently on near-optimal arms, improving the resulting decisions made under uncertainty.
We demonstrate these properties by examining seventeen large-scale experiments conducted on Facebook from April to June 2017.
\end{abstract}


\section{Introduction}

The routine use of online experiments at Internet firms motivates the development of tools and methods which are robust and reliable when used for thousands of experiments per day, either through the use of manual self-serve tools by non-experts, or automated experimentation techniques like multi-armed bandit optimization. 
These experiments commonly have many variants or \emph{arms}; this often leads to the observed best performing arm being an overestimate of the true effect when using standard maximum likelihood estimation~\citep{gelman2012we}.
This behavior, then, can lead to suboptimal inferences about which arm is best.
In this paper, we present a simple and more accurate estimation tool--- shrinkage---which is particularly well suited to routinized digital experimentation and which leads to better resulting decisions by hedging more efficiently when the best arm is uncertain. 
It is easy to understand, effective, and perhaps most importantly, it is highly robust.
This makes it readily able to serve as a default tool in experimental analyses.

A conception of ``scalability'' is the distinguishing feature of online experimentation. 
In this context, that implies a number of particular features: (i) sample sizes range from hundreds to hundreds of millions (ii) the number of treatment groups in an experiment range from a few to hundreds (iii) many practitioners will have no expert training (e.g. they are software engineers, not statisticians) (iv) many business-relevant effects are small in magnitude (v) most experiments are run with a view towards optimization~\citep{letham2018}\footnote{This paper will deal with more traditional experimentation under SUTVA~\citep{imbens2015causal}, rather than difficult to study phenomena such as peer effects~\citep{bakshy2012,eckles2016}}.

This work provides an estimator of causal effects in experiments which aligns with these principles. Respectively, it (i) requires only summary statistics (means and variances) which are already, by necessity, collected for experiments \footnote{This is unlike covariate adjustment~\citep{bloniarz2016,lin2013} which requires additional infrastructure and domain knowledge to implement well~\citep{deng2013improving}.}, (ii) provides larger benefits as the number of similar treatments are increased, (iii) is easy to implement and understand, (iv) increases the accuracy of estimated effects, allowing less uncertainty for the same size of experiment, and (v) finds the best arm faster and more effectively.

The optimization-based perspective on experimentation is not isolated to technology firms, but is also common more broadly within the social sciences~\citep{imai2010,benartzi2017,madrian2014} and elsewhere~\citep{zhang2012robust}. 
Whenever there is agreement within a group or organization about what outcomes should be improved, experimentation can be used as a tool to achieve these ends, whether those outcomes are revenue, voting behavior, or public health. 

This paper will proceed by first laying out the proposed shrinkage estimator (a standard James-Stein estimator) for the estimation of causal effects as well as an appropriate variance estimator (not previously analyzed in depth in the extant literature). 
We demonstrate the consistency of these estimators and then proceed to examine various properties of them. 
Next, we introduce a series of seventeen experiments run on Facebook from April to June 2017. 
This experimental data will be used to examine the performance of the shrinkage estimators on representative data. 
This performance will be evaluated in two broad ways. 
First, we will evaluate the performance and properties of the shrinkage estimator in the context of one-shot experimentation, addressing such questions as its accuracy and its coverage. 
We will conclude with an evaluation of its performance in sequential experimentation, demonstrating its effectiveness in optimization, a less well-understood application of empirical Bayes.
We find that shrinkage estimation makes sequential decision-making more robust by efficiently exploring arms in the vicinity of the optima.
This thereby increases the likelihood that the best arm will be found and reduces the regret accrued by experimentation, resulting in better sequential decision making.
 
\section{Shrinkage}
Our setting for the remainder of the paper is that of a randomized control trial, performed online or otherwise.
We consider the experiment in which we observe data from $K$ treatment groups (used interchangeably with `arm'). 
Our inferential target will be the average response in each group.
For each group, we only observe the sample mean, $m_k = \frac{1}{n_k} \sum_{i:d_i=k} y_i$ where $d_i$ is the $i$th unit's treatment assignment, $n_k$ is the number of units assigned to the $k$th treatment group and $y_i$ is the $i$th unit's observed outcome.
We also observe the standard error of this quantity.
Since we assume a simple setting with randomization and no interference / spillovers, the sample means are unbiased for our target parameters, $\mu_k = \mathbb{E}[y_i(k)]$ \citep{imbens2015causal}.
Our goal is to take these highly aggregated quantities and construct an estimator better than $m_k$ by using the fact that each of these true effects share some underlying common distribution with a central tendency.
That is, we wish to improve our estimation without using any additional, difficult to acquire unit-level information.
We will do this using shrinkage.

To motivate our shrinkage estimator, we assume that the observed sample means $m_k$ are drawn from a normal distribution centered at $\mu_k$ with variation $\sigma^2$. 
That is, we assume that the sampling distribution of $m_k$ is actually normal. 
Of course, this is a fairly weak assumption, as we know that the Central Limit Theorem guarantees that this will be true asymptotically. 
That is, we are simply assuming that our sample size is large enough that the Central Limit Theorem holds. 
Note also that we are assuming homoskedasticity amongst our various $m_k$. 
Similar derivations are straightforward in the case of variances that differ by treatment group, but those will not be addressed here.

The problem of estimating $K$ means, $m_k$, is very familiar to students of statistics. 
\citep{stein1956} shows that the aggregate accuracy (e.g. compound mean squared error) of the $m_k$ for $K > 3$ are inadmissable for estimating the underlying means $\mu_k$ of a set of normally distributed variables. 
That is, there is guaranteed to be a better estimator of $\mu_k$ than the maximum likelihood-based approach.

In a certain sense, this isn't surprising -- one of the reasons we analyze experiments with the sample means is because it allows for \emph{unbiased} estimation of causal effects. 
This formulation, however, clearly demonstrates that while these effects are efficient among the class of unbiased estimators, they are not best possible estimators if we broaden our scope to include estimators which admit small amounts of bias in exchange for greater overall accuracy.

In cases like online experimentation, trading off small, well-understood biases for greater accuracy is likely a trade worth making. 
Thus, we motivate an improved estimator which is guaranteed to be both consistent and more accurate than the na\"ive estimator.

In particular, the estimator we will focus our attention on is the positive-part James-Stein estimator~\citep{efron1971,efron1975}:

\begin{equation}
\label{ass:normal}
m_k^{JS} = \bar{m} + (1 - \xi_k) (m_k - \bar{m})
\end{equation}

where $\bar{m}$ is the mean of all $m_k$, $\sigma_k$ is the standard error of the means and where
$$
\xi_k = \text{min}\left(\sigma_k^2 \frac{K-3}{s^2}, 1\right)
$$,

where $s^2 = \sum_{k=1}^K (m_k - \bar{m})^2$. We replicate a standard derivation of this estimator in appendix~\ref{app:bayes_basic}.

With a fixed number of arms, fixed $\mu_k$ and increasing sample size (allocated with nonzero probability to each arm), then $\frac{K-3}{s^2}$ will converge to a constant, while each $\sigma_k^2$ will asymptotically approach zero. 
Thus, $\xi_k$ (the amount of shrinkage) will approach zero by Slutsky's theorem. This implies that the estimator will converge to the same limit as the unbiased MLE. 
In finite samples, some amount of shrinkage will be performed so as to increase accuracy, but this will be reduced as samples increase.

Typical treatments of James-Stein estimators treat $\bar{m}$ and $\frac{s^2}{K-3}$ as fixed and known a priori~\citep{efron2012}. 
Thus, the most commonly used variance estimator for equation~\ref{ass:normal} would be, simply:

\begin{equation}
\label{eq:varbasic}
\mathbb{V}(m_k^{JS}) = (1-\xi_k) \sigma_k^2
\end{equation}

This shows that with a known common mean, as the amount of shrinkage (indicated by $\xi$) increases, the variance mechanically decreases to zero. 
In our setup, however, these quantities are not known, but estimated, so it is extremely undesirable for the variance to converge to zero, as it will lead to strongly anti-conservative inference. 
We can thus construct a variance estimator which incorporates the necessary additional sources of estimation error as follows:

\begin{equation}
\label{eq:varbetter}
\mathbb{V}(m_k^{JS}) \approx (1-\xi_k) \sigma_k^2 + \frac{\xi_k s^2}{K} + \frac{2 \xi_k^2 (m_k - \bar{m})^2}{K-3}
\end{equation}

There are a few things to note about this expression. 
First, the first term is identical to the variance estimator in equation~\ref{eq:varbasic} that takes the common distribution as known. 
This term represents the contribution of the uncertainty in the estimation of $m_k$. 
The second term incorporates the uncertainty from the estimation of $\bar{m}$. 
The final term adds in the uncertainty from the dispersion of the effects. 
We derive this expression for the variance from Bayesian foundations with uninformative priors in appendix~\ref{app:bayes_full}.

The most clear objection to these formulations of a James-Stein estimator is that it will perform less well when the distribution of effects are non-normal. 
For instance, if effects are mostly near zero with some outlying 'true' effects, then simple James-Stein will tend to perform poorly. 
This motivates the use of a limited translation variant (a la \citet{ghosh2009}) of the James-Stein estimator which reduces or eliminates shrinkage for outlying arms. 
In what follows, we prefer to be more conservative in our estimates for extreme arms in order to best insulate us from excessive optimism in the face of uncertainty.
Thus, the remainder of this paper will focus on the benefits of the estimators in equations \ref{ass:normal} and \ref{eq:varbetter}.

\section{Experiments}

\begin{figure}
  \begin{center}
    \begin{subfigure}[t]{0.225\textwidth}
        \begin{center}
        \fbox{\includegraphics[width=.8\textwidth]{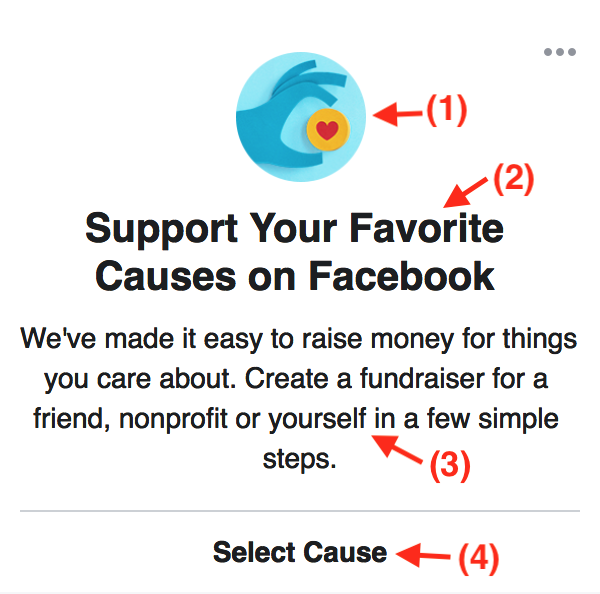}}
        \end{center}
        \caption{Example messaging experimental condition. Factors in the factorial design are elements like image (1), title (2), text (3), and button text (4).}
        \label{fig:example_condition}
    \end{subfigure}
    \begin{subfigure}[t]{.225\textwidth}
      \begin{center}
      	\includegraphics[width=.9\textwidth]{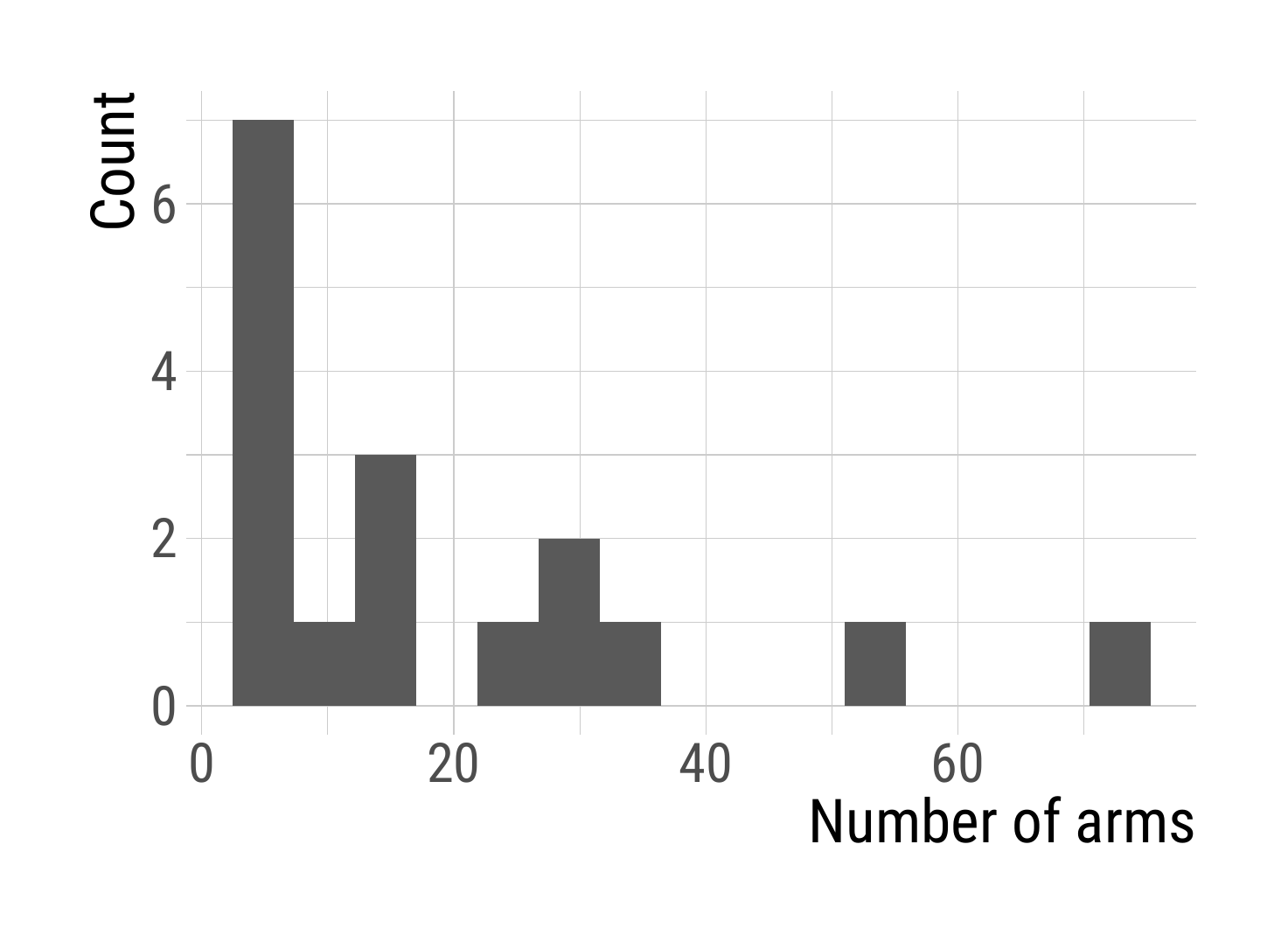}
      \end{center}
      \caption{\centering Number of arms in each experiment.}
      \label{fig:num_arms}
    \end{subfigure}
    \begin{subfigure}[t]{.225\textwidth}
      \begin{center}
      	\includegraphics[width=.9\textwidth]{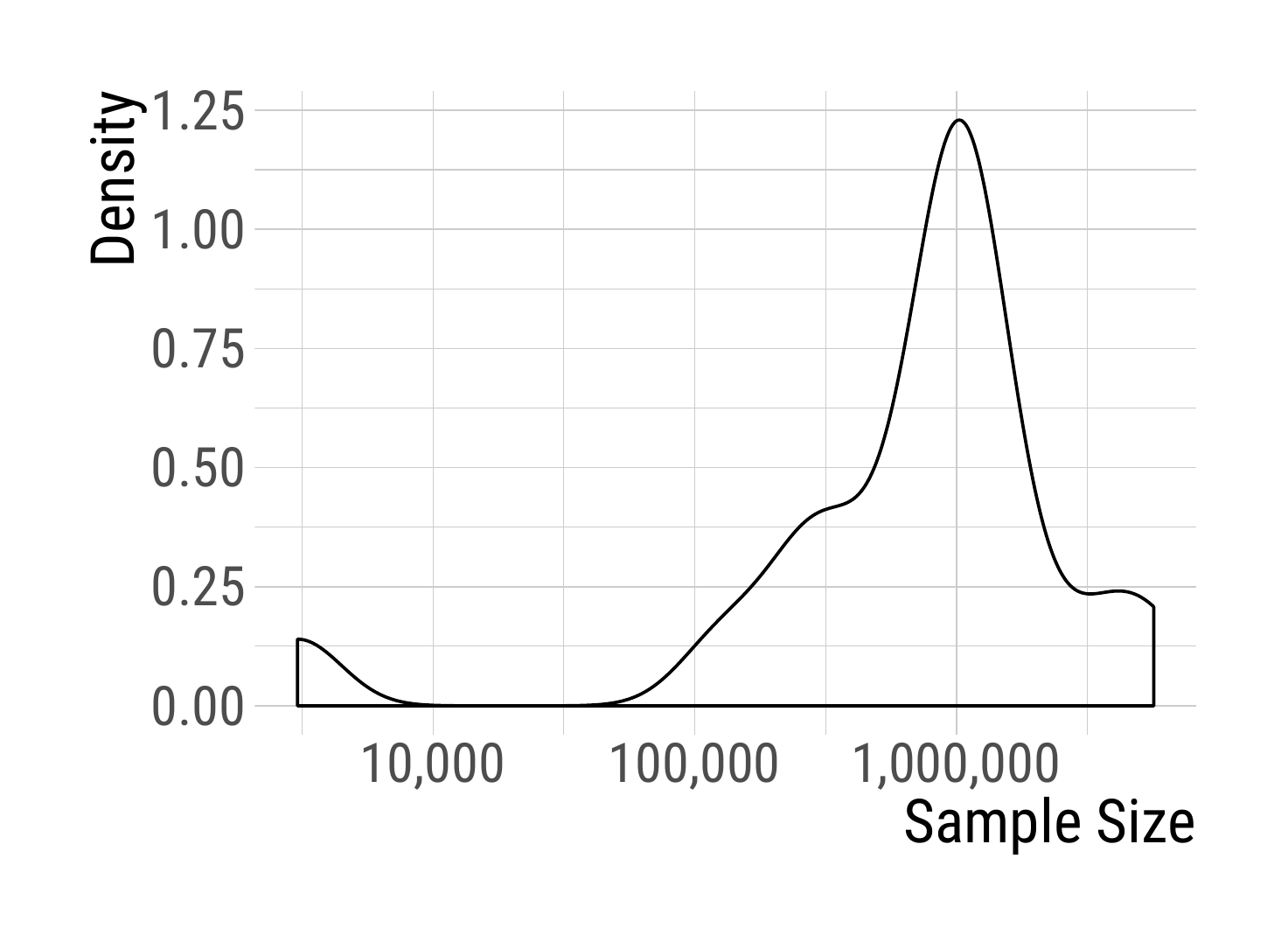}
      \end{center}
      \caption{\centering Kernel density estimate of sample size in each experiment.}
      \label{fig:num_users}
    \end{subfigure}
    \begin{subfigure}[t]{.225\textwidth}
      \begin{center}
      	\includegraphics[width=.9\textwidth]{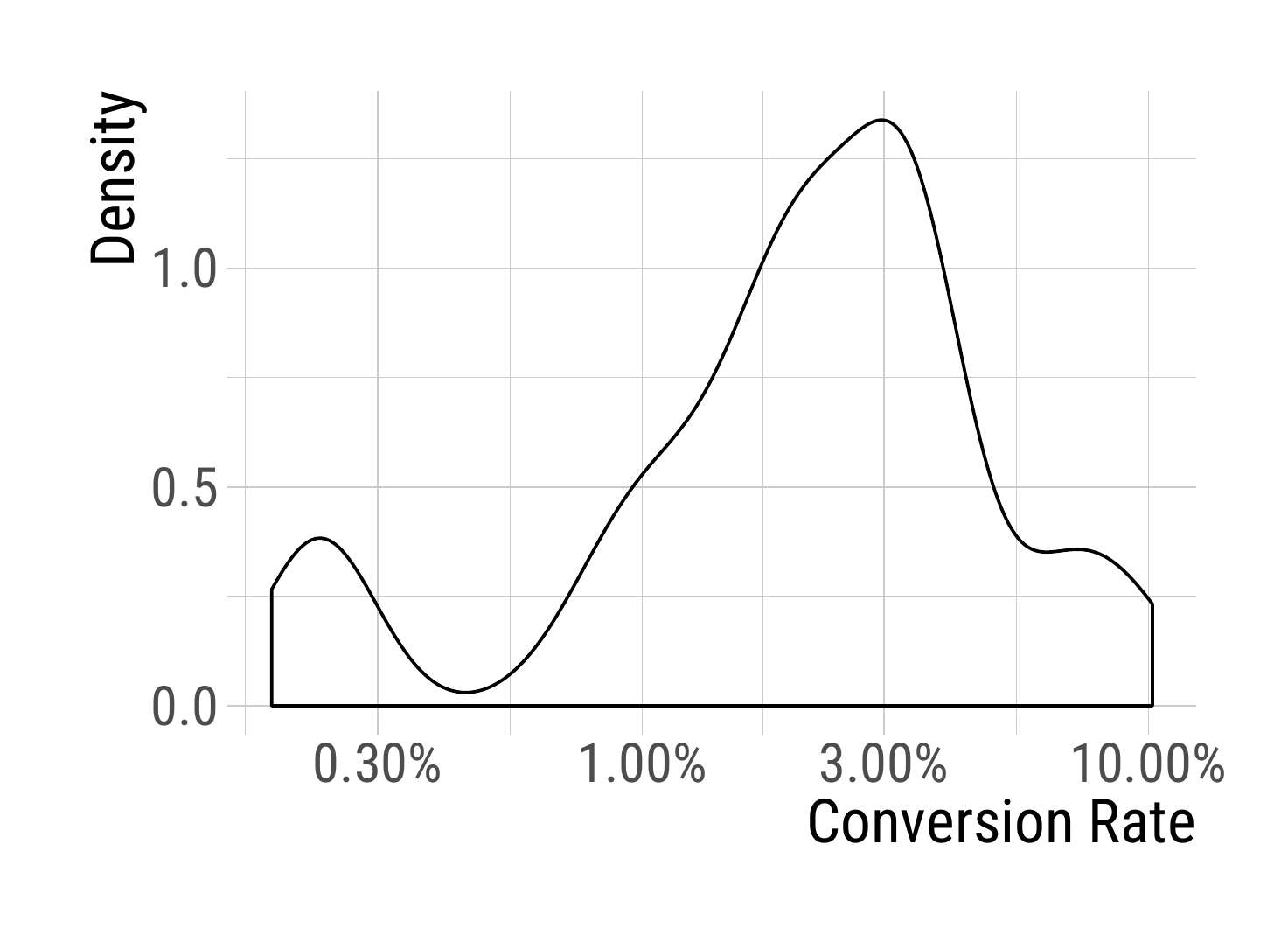}
      \end{center}
      \caption{\centering Kernel density estimate of conversion rate (for each arm and experiment), weighted by sample size.}
      \label{fig:conversion}
    \end{subfigure}
  \end{center}
  \caption{Summary information on the experiments used in this study.}
\end{figure}

In the simulation studies that follow, we take the means and variances from seventeen recent routine online experiments conducted on Facebook intended to optimize content in the normal course of operation of the platform. 
These experiments appear at the top of News Feed as in figure \ref{fig:example_condition}. 
These experiments were full-factorial \citep{box2005statistics} experiments testing different factors like wording, image, button text and layout of ``calls to action''. 
That is, each of the factors were independently randomized, generating an arm for every possible combination of (for instance) wording and image.
In general, the substance of these experiments is mundane: encouraging users to upgrade their app versions, update their email addresses or (as in figure \ref{fig:example_condition}) donate to a charity of one's choice. 
Treatments were essentially persuasion strategies for inducing individuals to take an action (the action varied by experiment), and the outcome was whether the user took the desired action or not (called a conversion). 
In the example in figure \ref{fig:example_condition}, the action would be whether the user chose to create a charitable fundraiser or not. 

The number of arms in these experiments varied from a minimum of 3 treatment groups to a maximum of 72 groups (see Figure~\ref{fig:num_arms}). 
The sample sizes varied from 3000 to over 5 million (see~\ref{fig:num_users}). 
Combined, these experiments consisted of over 20 million individuals. 
The typical conversion rates were less than 10\%, with the sample size weighted average being approximately 2.5\% (see~\ref{fig:conversion}). 
Given the factorial structure (and relative unlikelihood of there being strongly interactive effects), we wouldn't expect any single arm to be vastly superior to all other arms. 
As such, we would expect there to be at least some central tendency of effect sizes and, therefore, shrinkage to be effective.

Our evaluations focus on three dimensions: accuracy (measured by mean squared error), frequentist coverage and sequential decision making (measured by frequency that the best arm is played and regret).

\subsection{Static Simulations}
In the simulations that follow, we will treat the estimates from these experiments as the ground truth. 
Our simulation studies will redraw data from a (Normal) parametric bootstrap downsampled to a significantly smaller sample size (20\%). 
Note, however, that the true distribution of effects will still be non-normal.
This will generate similar data to that which we encounter in our day-to-day experimentation at Facebook. 
The performance of our methods on this data, then, will do a good job of indicating how well we will do when we apply them to future experiments.

\begin{figure}
\begin{center}
\includegraphics[width=.65\textwidth]{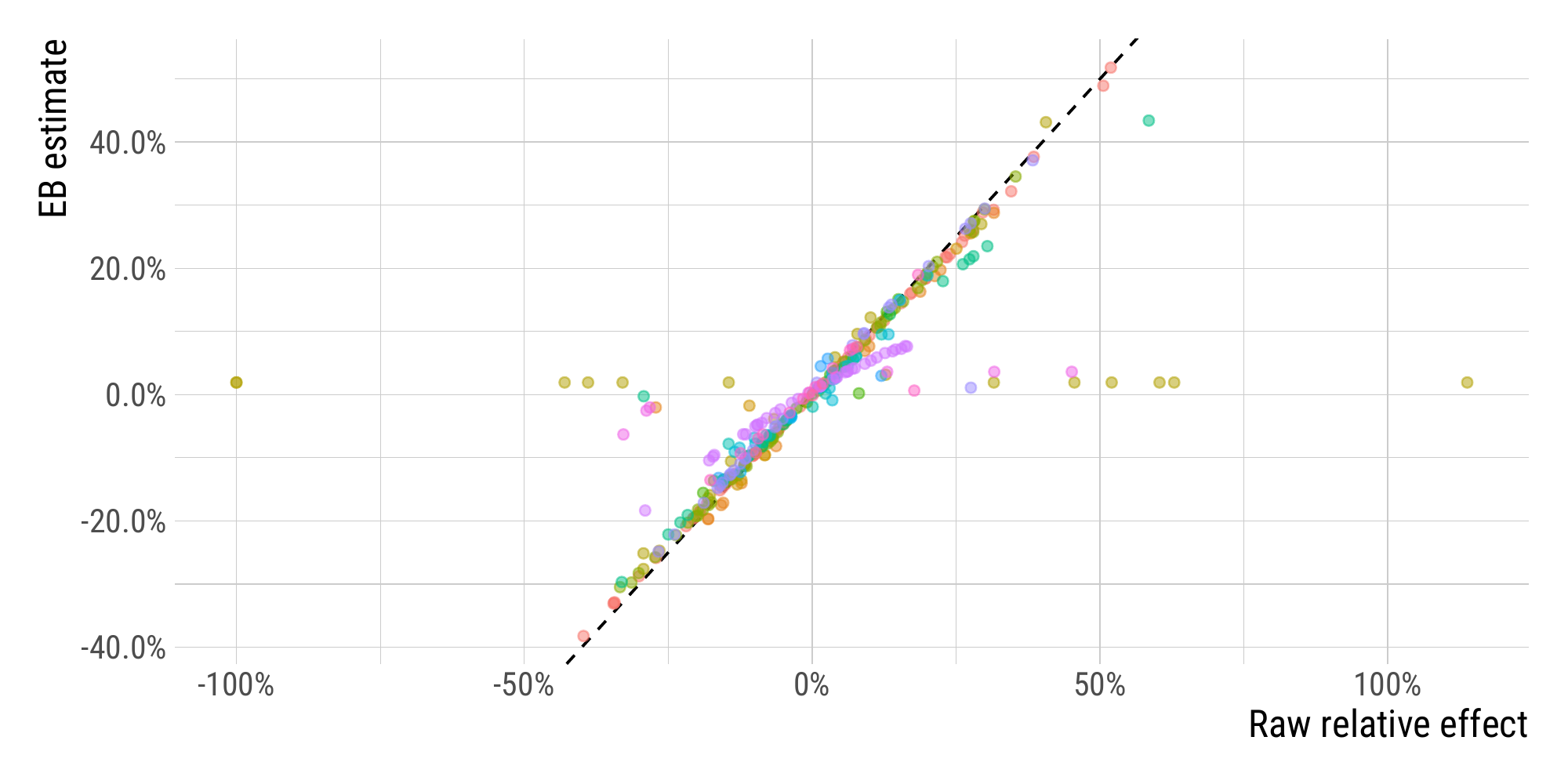}
\end{center}
\caption{Empirical Bayes shrinks extreme arms the most. Each point is a single point estimate of one arm from a single experiment. Arms lying near the dashed line have less shrinkage than those far from the line. Displayed effects are relative to the overall average within the experiment. Colors indicate experiments.}
\label{fig:shrinkage_by_arm}
\end{figure}

The first thing to note is the amount of shrinkage that is performed on each arm in Figure~\ref{fig:shrinkage_by_arm}. 
All else equal, the most extreme values are shrunk the most towards the middle of the distribution (note that the sample sizes vary by arm, leading to some non-linearities). 
This is an extremely desirable property. 
Outlying values are, in general, likely to be over or under estimates of the true effect. 
This is true through the same mechanism as the statistical significance filter: conditioning on observing a treatment as the 'best' among a set of treatments, it is more likely to be an overestimate of the true effect. 
For example, it may be readily observed that the expectation of the maximum of two identical Normal distributions is larger than the expectation of either~\citep{nadarajah2008}. 
In practice, particularly when our experimental objective is optimization, we focus our attention on the extreme arms, but more important than estimating the magnitude of this effect with minimal error, is discovering \emph{which arm is best}. 
By concentrating shrinkage on these extreme arms, we guard against the multiple comparisons issue omnipresent in large online experiments. 
This is true for the same reason that estimates from multilevel models do not tend to suffer from the same problems around multiple comparisons as do the maximum likelihood estimates~\citep{gelman2012}.

\begin{figure}
\begin{center}
\includegraphics[width=.6\textwidth]{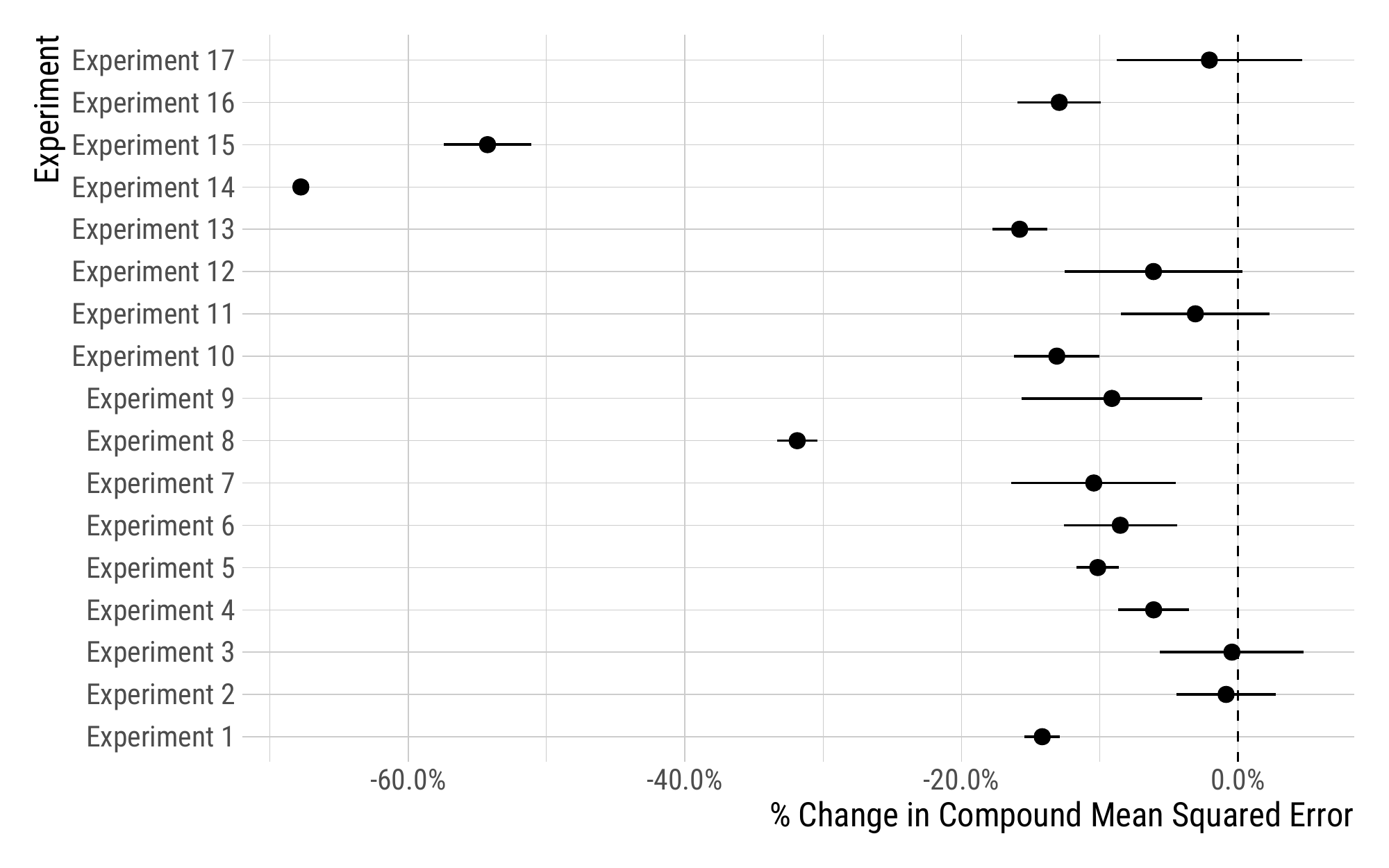}
\end{center}
\caption{All experiments gain accuracy from empirical Bayes. This plot shows the change in compound mean squared error for each experiment attained by switching from the MLEs to the empirical Bayes estimates. Standard errors of the ratio are calculated using the Delta method.}
\label{fig:mse_overall}
\end{figure}

The overall performance of the method in reducing compound mean squared error (that is, the sum, for each experiment, of the MSE of the arms) can be seen in Figure~\ref{fig:mse_overall}. 
The MSE does not increase for any experimentation under examination. 
Indeed, most see around a 10\% reduction in MSE (with an average reduction of around 15\%). 
Some experiments, however, see significantly greater improvements in accuracy; experiments 14 \& 15, for instance, reduce MSE on the order of 50\%.
These results match exactly the sense in which a James-Stein estimator renders the MLE inadmissable.

\begin{figure}
    \centering
    \includegraphics[width=.6\textwidth]{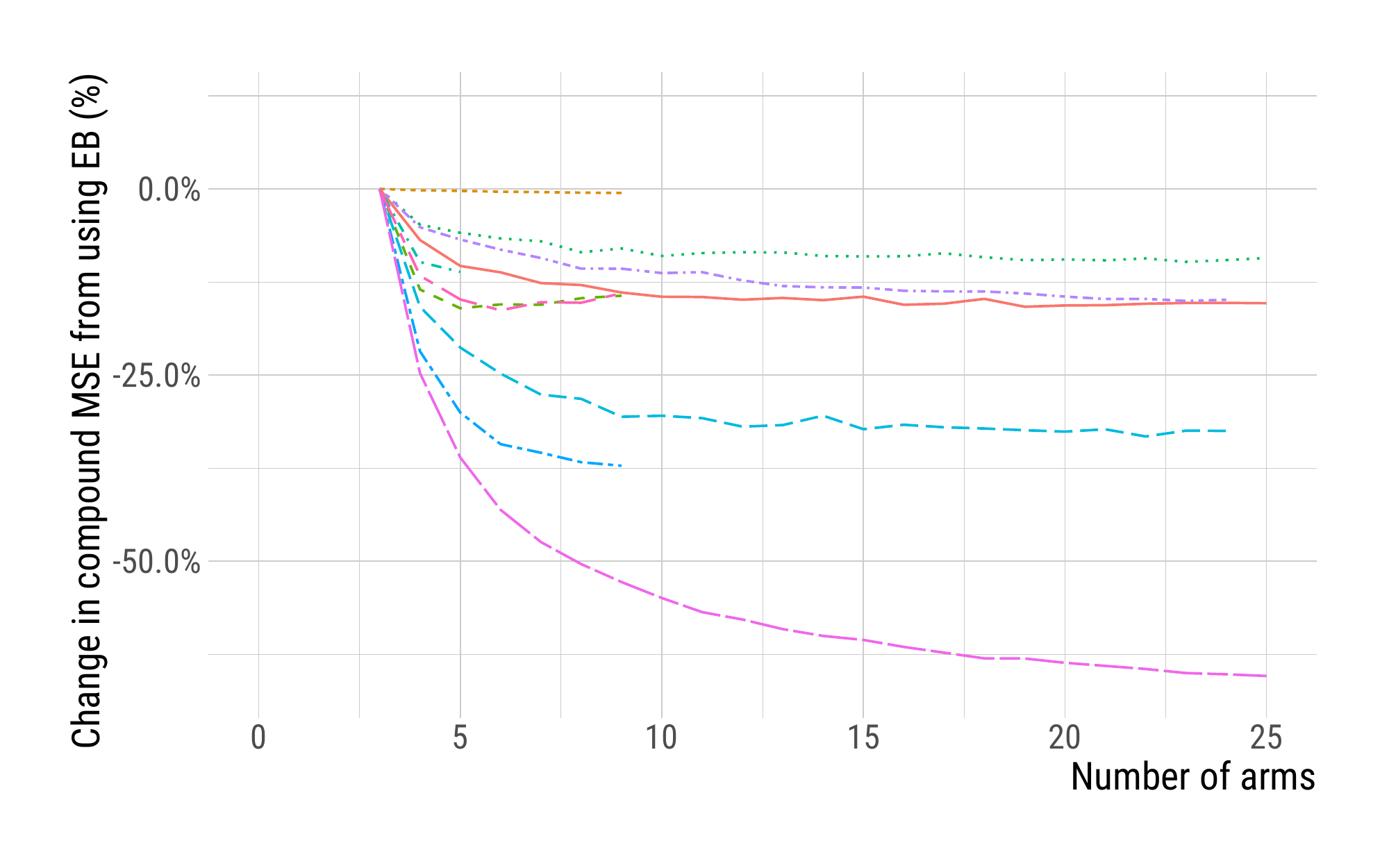}
    \caption{Empirical Bayes confers benefits even with relatively few arms. The relative gain in compound MSE varies as a function of the number of arms. Each line represents a series of simulations with a variable number of arms derived a single real-world experiment.}
    \label{fig:mse_by_num_arms}
\end{figure}

Figure~\ref{fig:mse_by_num_arms} demonstrates how the accuracy enhancing benefits accrue as a function of the number of arms. 
To generate this figure, we simply subsampled (without replacement) a number of the original arms in the experiment and calculated compound MSE as above. 
This figure demonstrates that a large number of arms are not necessary for empirical Bayes to substantially improve accuracy. 
Indeed, so long as there are three arms empirical Bayes provides greater accuracy than the MLEs, and by around 10 groups, the gains are nearly as large as they'll ever be. 

Clearly, if the goal of estimation is to reduce the overall accuracy across over all arms, shrinkage estimation is to be greatly preferred over the MLE. 
But in fact, there are stronger statements we can make.
85\% of arms see a reduction in MSE from using empirical Bayes -- therefore, the \emph{modal} and \emph{median} arm also see gains.
Such reductions in MSE are of particular importance when seeking to understand tradeoffs between multiple metrics.
In such cases, it's important to have as much accuracy as possible for as many arms as possible.

\begin{figure}
\begin{center}
\includegraphics[width=.6\textwidth]{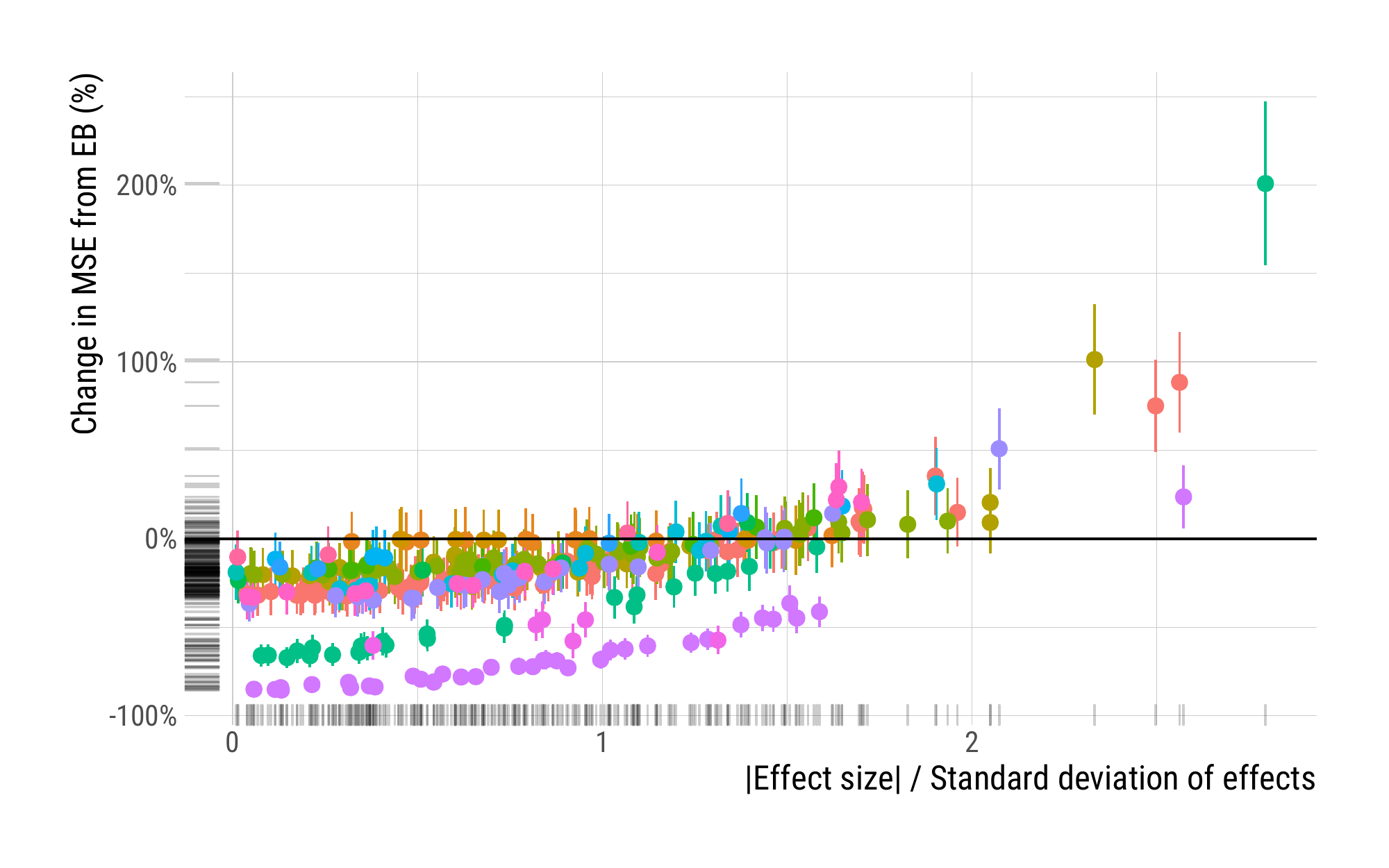}
\end{center}
\caption{The majority of arms are estimated more accurately with empirical Bayes. Each point shows the relative change in MSE from switching from the MLE to EB for a single arm in one experiment. The x-axis is the ratio of the arm's effect size divided by the standard deviation of effect sizes in its experiment. Colored by experiment.}
\label{fig:mse_by_arm}
\end{figure}

Nevertheless, a method which provides small improvements in MSE for most arms, but a huge loss in MSE for the most important arms might not be a reasonable tradeoff to make. 
Figure~\ref{fig:mse_by_arm} shows the change in MSE from using empirical Bayes for each arm in each experiment. 
What can be seen is that only extreme arms tend to suffer in terms of MSE, while arms toward the center of the distribution tend to reap substantial benefits. 
Overall, around 90\% of arms improve their MSE from using empirical Bayes.
Of course, this increase in MSE is also exactly why we are better insulated against type M errors (overestimation of effect size)--- estimates are brought more in line with the central tendency. 
This causes us to erroneously believe that they are more similar to other arms than they are. 
In practice, the desirability of this tradeoff will be based on how it affects the decisions made, analyzed in section~\ref{sec:dynamic}.

For some experiments, (like experiment 14) there is no real tradeoff -- every arm has lower MSE than the average MSE from the MLE. 
That said, other experiments (such as experiment 13) have both arms with substantially lower MSE and some with higher MSE. 

\begin{figure}
\begin{center}
\includegraphics[width=.6\textwidth]{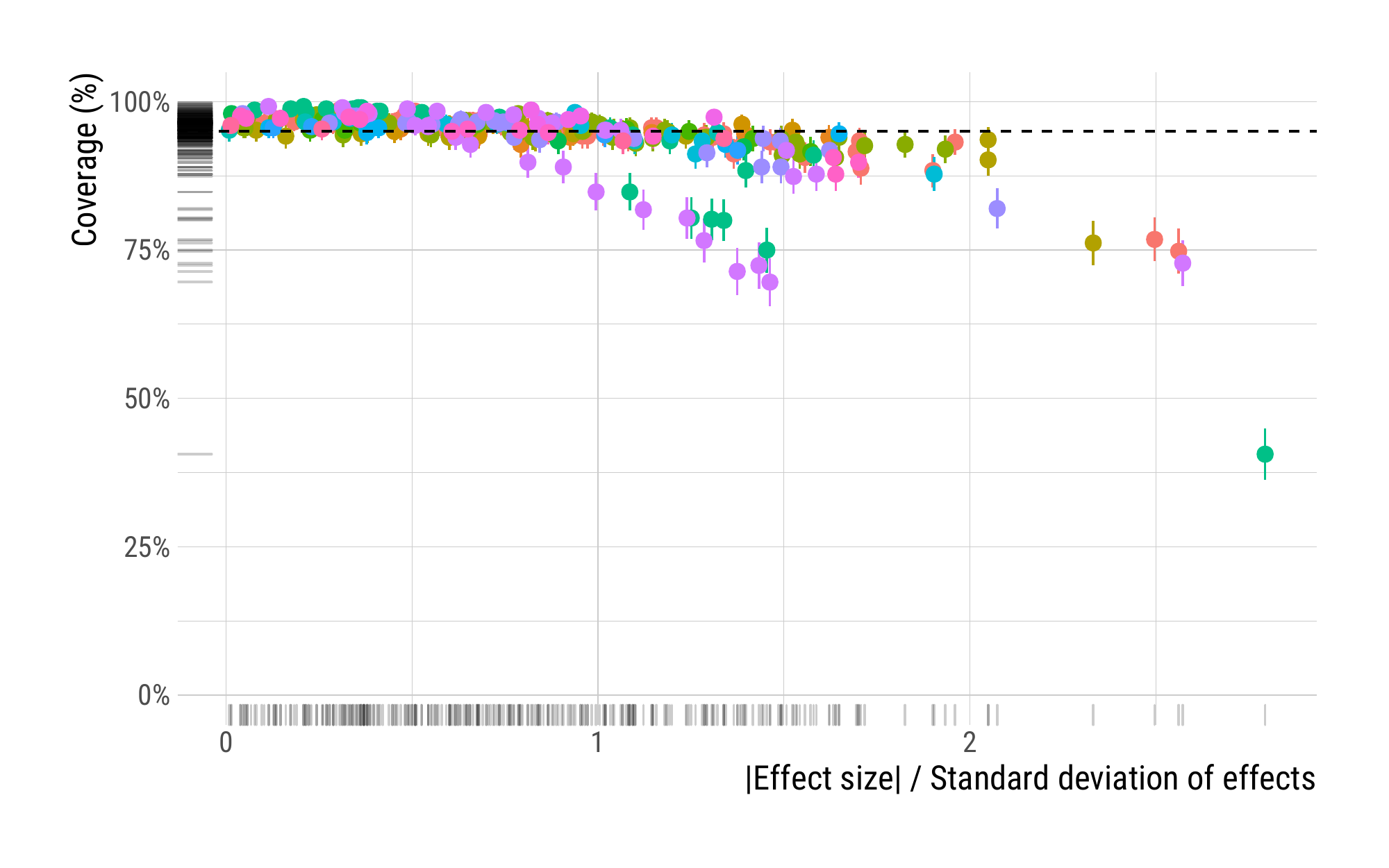}
\end{center}
\caption{Most arms retain strong frequentist coverage through the use of empirical Bayes. Each point is a single arm in one experiment. The realized coverage for each arm based off of a nominal 95\% confidence interval. The x-axis is the ratio of the arm's effect size divided by the standard deviation of effect sizes in its experiment. Colored by experiment.}
\label{fig:coverage_by_arm}
\end{figure}

Point estimates are not everything; we also want to be sure that our measures of uncertainty are appropriate, particularly given the improved variance estimator we have provided in equation~\ref{eq:varbetter}. 
We examine this through the frequentist properties of the shrunk estimates. 
We examine the coverage (the proportion of time our 95\% confidence intervals contain the ground truth) attained for the estimates of individual arms, displayed in Figure~\ref{fig:coverage_by_arm}.

We can see that most arms actually have higher than nominal coverage (60\% of points have higher than nominal coverage, and 90\% have at least 90\% coverage). 
By comparing Figure \ref{fig:mse_by_arm} to figure \ref{fig:coverage_by_arm}, we can see that the same extreme arms which received the most shrinkage tend to suffer most in terms of both MSE and coverage.
These are the arms which move their means most through shrinkage, so it is unsurprising that they suffer in terms of their frequentist properties. 
That said, however, this reduction in coverage is not entirely bad. 
We are particularly concerned with reducing the risk of type M errors, in which we estimate the effects of an arm too optimistically~\citep{gelman2014beyond}. 
The reduced coverage we observe is intricately connected with the results observed in the following section.

\subsection{Dynamic Simulations}
\label{sec:dynamic}

In this section, we seek to validate the performance of shrinkage estimation when combined with sequential experimentation. 
First, an aside on our methodology for performing sequential experimentation. 
The method we use for sequential experimentation is batch-based Thompson sampling~\citep{scott2010}. 
Thompson sampling is a common means of navigating the exploration-exploitation tradeoff in practice which is both simple and easy to understand: at each step, we assign units proportionally to our posterior that they are the best.
The process we follow, is to first run a large ``exploration'' experiment with a wide array of different arms (for instance, a large full-factorial experiment). 
After this phase, we perform Thompson sampling to choose the appropriate arms for the next phase.
This can be estimated simply by drawing samples from the posterior distribution of all arms. 
That is, we take each arm's mean as distributed independent normal by the CLT. 
Thompson sampling then amounts to sampling a large number of times from $K$ independent normals. 
We build a posterior by recording, for each set of $K$ samples, which arm's sample was largest. 
The distribution of maximal arms, then, is the distribution over treatment assignment to be used in the following batch of the experiment. 
In addition to working well in practice~\citep{chapelle2011}, Thompson sampling has more recently been shown to have good theoretical properties as well for minimizing discrepancy in the outcome between the best possible arm and the chosen arms in a sequential experiment~\citep{agrawal2012,agrawal2013}.

In these simulations, because we can't assume that the normality assumption will hold (due to the small conversion rates relative to sample size), we will form our posterior for Thompson Sampling using beta distributions. 
After each batch, that is, we will fit a beta distribution to the set of $m_k$ in the simulation, and add in the estimated shape parameters to the observed successes and failures for each arm (which make up the shape parameters under maximum likelihood estimation). 
We then draw the posterior distribution of optimal arms from these Beta distributions with empirical Bayes priors.

\begin{figure}
\begin{center}
\includegraphics[width=.65\textwidth]{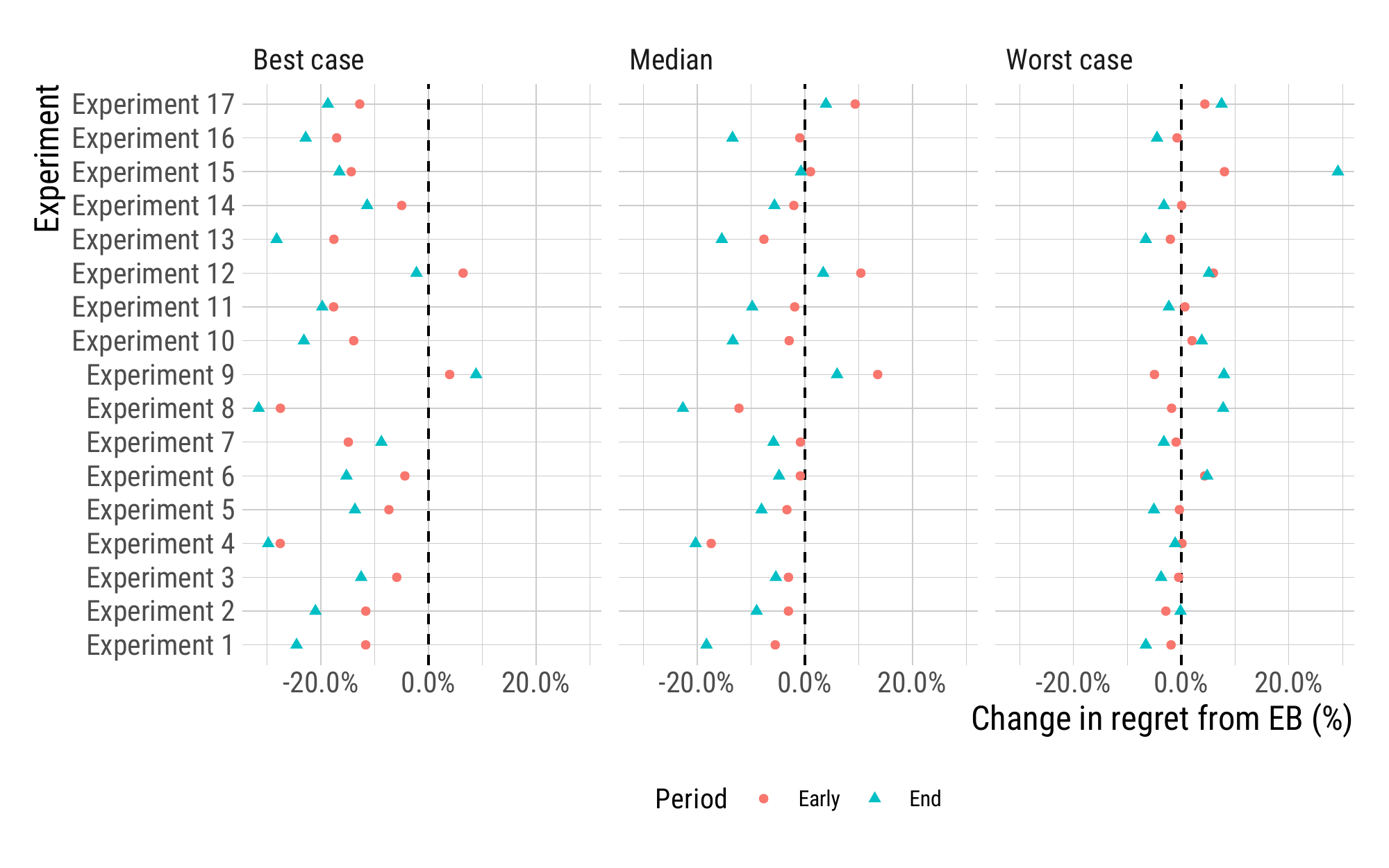}
\end{center}
\caption{Empirical Bayes shrinkage tends to improve best and median-case performance of bandit optimization. Each panel shows how cumulative regret changes when posteriors for Thompson sampling are computed using the empirical Bayes estimator, rather than the MLE.  Panels represent the 2.5th, 50th, and 97.5th percentile of cumulative regret over simulations from left to right. Lower regret is better. Comparisons are shown for both the end of the sequential experiment ("End") and 25\% through the experiment ("Early").}
\label{fig:ab_improve}
\end{figure}

Figure~\ref{fig:ab_improve} shows the improvement of empirical Bayes relative to MLE for constructing a posterior. 
That is, the figure plots how (over 500 simulations) cumulative regret (the cumulative difference between the best possible reward and the realized reward) changes by using empirical Bayes. 
In Figure~\ref{fig:ab_improve}, the three panels demonstrate, the performance in a `best case' at the 2.5 percentile (over simulations), the median case and a `worst case' at the 97.5 percentile.
Moreover, this figure presents results for two points in time: early in the experiment (25\% through the procedure) and at the end of the experiment.
In the far left panel, it is clear that empirical Bayes provides substantial gains on the order of 10\% to 20\%  by the end of the experiment.
In terms of median performance, empirical Bayes results in typical gains of around 10\% at the end of the experiment.
These gains tend to be smaller (but nonzero) early in the experiment, with empirical Bayes accruing fairly similar median regret than the MLEs as each method concentrates on exploration.
Together, these two results are indicative of a more efficient exploration strategy by empirical Bayes which results in a greater chance of successfully finding the optimal arm to play.
In the ``worst case'' shown in the panel on the right, empirical Bayes has relatively similar performance to the MLE.
Crucially, across all three panels there is not a clear example in which the MLE clearly outperforms empirical Bayes.
At worst, in the three cases where the median-case performance of empirical Bayes is a few percentage points below that of the MLE at the end of the experiment.
In two of these cases, empirical Bayes still outperforms the MLE in the best-case analysis.
In short, empirical Bayes improves the typical and the best case performance of sequential optimization.

\begin{figure}
\begin{center}
\includegraphics[width=.65\textwidth]{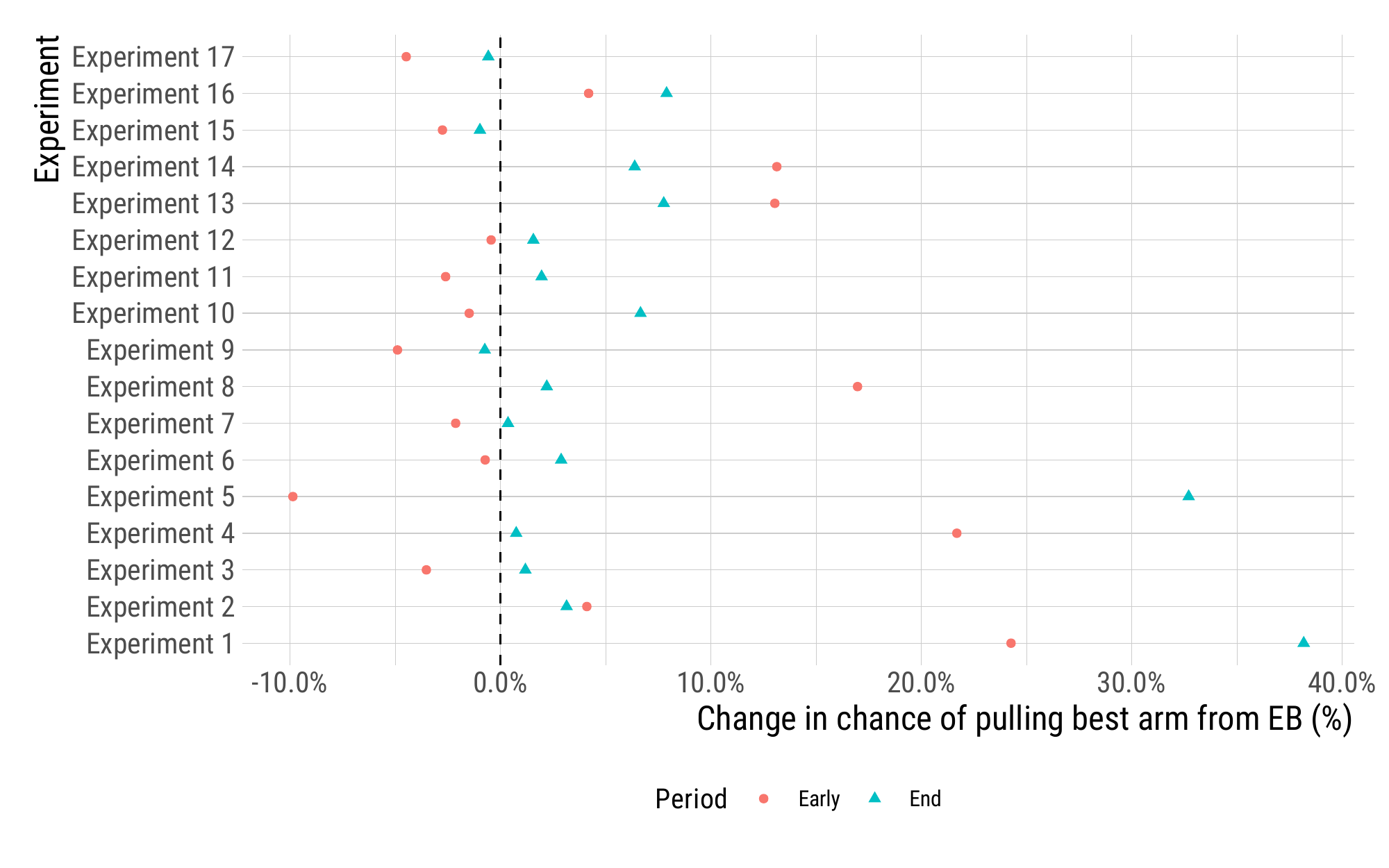}
\end{center}
\caption{Empirical Bayes is more likely to play the best arm, particularly in later rounds. Displayed is the probability that the sampling algorithm chooses the best arm for early in the experiment and at the end of the experiment.}
\label{fig:best_arm}
\end{figure}

The same general trends may be observed along a related metric: allocating units to the best arm~\citep{audibert2010}. 
Figure~\ref{fig:best_arm} shows how often Thompson sampling with MLEs and empirical Bayes estimates allocate users into the best treatment group. 
Shrinkage is able to ensure that the best arm is played more often than relying on the MLEs alone. 
While early in the experiment, empirical Bayes often plays the best arm less often than one would when using the MLEs, by the end of the experiment there are substantial gains.
See, for instance, how in experiments 1 and 5, Thompson sampling on the empirical Bayes estimates provide as much as a 30\% to 40\% increase in the chance of allocating units to the best arm. 
For no experiment does empirical Bayes play the best arm less often -- at worst, the use of empirical Bayes performs comparably to the MLEs.
The typical increase in playing the best arm is between 5\% and 10\%.

A comparison to the results on coverage in the previous section (e.g. Figure~\ref{fig:coverage_by_arm}) demonstrates an important fact: it's exactly the experiments with arms that have below-nominal coverage which attain substantial gains in sequential optimization. 
In other words, the under-coverage belies the fact that shrinkage \emph{makes better decisions about the best arm} when used within Thompson sampling.
Since our whole exercise is aimed at improving our statistical decisions, this is resoundingly a tradeoff worth making.

\begin{figure}
\begin{center}
\includegraphics[width=.5\textwidth]{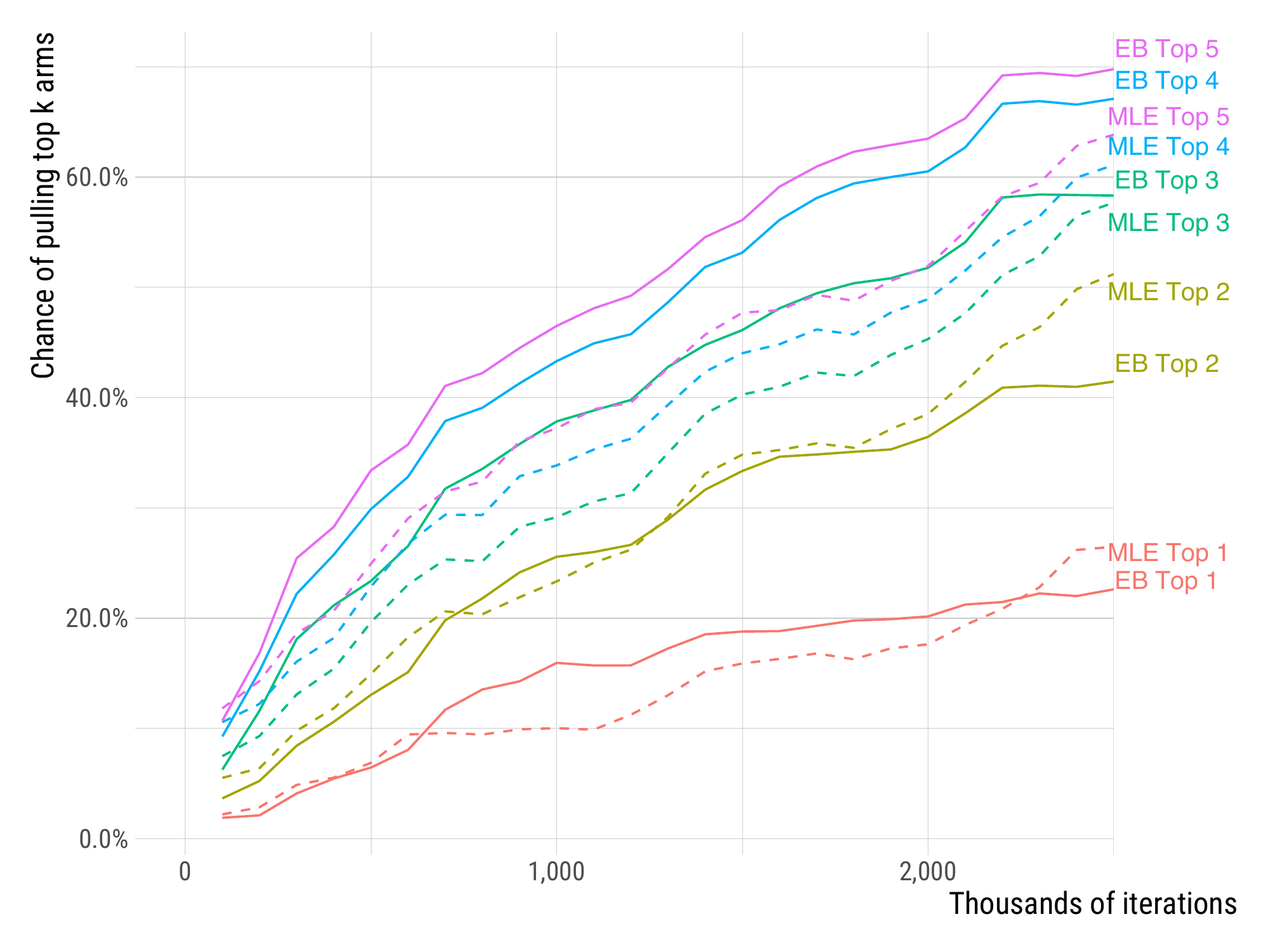}
\end{center}
\caption{Empirical Bayes concentrates exploration on the \emph{set} of the best arms. Displayed is the probability of sampling the top 6 arms as a function of time averaged over 50 simulations. Compared are Thompson sampling using a posterior constructed from the MLEs and from the empirical Bayes estimator.}
\label{fig:pr_top_k}
\end{figure}

It's easy to see why empirical Bayes out-performs the MLEs from Figure~\ref{fig:pr_top_k}.
In this simulation, Experiment 1 was taken as the ground truth, and samples were drawn in batches of 1000 according to Thompson Sampling over 50 simulation iterations. 
The proportion of the time each of the top 6 arms were played is plotted.
The early exploration period is shown here.
The benefits in this period from empirical Bayes can be seen very easily as improved exploration among the set of best arms.
While the MLE typically does a good job of playing the very best arm, empirical Bayes does a much better job of concentrating play among the top \emph{four} arms.
This, in turn, means that smart shrinkage avoids some of the common difficulties in Thompson sampling around distinguishing among the very best arms.

In essence, then, Thompson sampling with empirical Bayes shrinkage behaves similarly to algorithms like ``top two Thompson sampling"~\citep{russo2016simple}.
In this algorithm, Thompson sampling incorporates information about both of the top two performing arms in its construction of a posterior of the probability of an arm lying in the top \emph{two} ranks.
This, in turn, results in greater ability to statistically distinguish the two best arms (and thus does better in terms of best arm identification by gaining greater confidence in which arm is best).
Empirical Bayes with standard Thompson sampling, however, chooses the number of top arms to distinguish in a data-adaptive way.
When there is no clear evidence for the superiority of an arm, shrinkage causes the algorithm to hedge more strongly among the \emph{set} of best arms, not merely the top two.

\section{Conclusion}

This paper has introduced an approach which increases the accuracy and reliability of online experimentation in both one-shot and sequential experimentation. 
It provides clear gains in accuracy, and it tends to do a better job of optimization and best-arm identification than do the MLEs. 
Furthermore, the method is easy to describe and implement with little technical debt (e.g., it works entirely with commonly available summary statistics). 
As such, it provides the basis for an agnostic approach to analyzing experimental data that is particularly well suited to the online environment.  
Because of this, the empirical Bayes estimators presented here are used as our default estimator for many-armed experiments and automated experimentation services, such as those used to optimize creatives like the ones featured in Figure~\ref{fig:example_condition}.

Interesting avenues of future work might attempt to provide similar methods which provide analogous guarantees while taking into account slightly more structure in the data. 
Two such opportunities present themselves. 
First, the shrinkage estimator examined here did not take into account the factorial structure observed in the data. A similar approach tailored towards factorial designs could shrink towards a simpler main-effects only model. 
Second, online experiments typically collect a wide variety of outcome variables. 
Constructing an estimator using a multi-variate shrinkage method like Curds-and-whey~\citep{breiman1997} would likely result in further gains by considering the correlation structure.

More broadly, this paper demonstrates how shrinkage / regularization can greatly benefit the performance of experimentation with very few drawbacks by reining in extreme effect estimates when they differ greatly from those observed in other arms. 
These benefits of regularization are well-understood in a machine-learning and regression context, but are underused in online experimentation, despite their clear applicability.

\section*{Acknowledgements}
This research conducted under the auspices of Adaptive Experimentation, a part of Facebook's Core Data Science team. We thank Steven Howard for feedback and for his help with the derivations in the Appendix.

%
\bibliographystyle{ACM-Reference-Format}
\bibliography{main}

%
\appendix
\section{Derivation of Normal-normal James-Stein}
\label{app:bayes_basic}

For all of the following derivations, we wish to estimate $K$ means $\mu_k \in \mathbb{R}$ using $K$ observations $m_k \in \mathbb{R}$ with $m_k \sim \mathcal{N}(\mu_k, \sigma^2)$ where we assume $\sigma^2$ is known.

The following is standard. We set up the Bayesian model:

\begin{align*}
\mu_k \mid \mu, \tau^2 &\sim \mathcal{N}\left(\mu, \tau^2 - \sigma^2\right)\\
m_k \mid \mu_k, \nu, \tau^2 &\sim \mathcal{N}\left(\mu_k, \sigma^2\right)
\end{align*}

with $\mu \in \mathbb{R},\tau^2 \in [\sigma^2,\infty)$ unknown. (The unusual parametrization of $\tau^2$ is convenient below.) 
The posterior distribution of $\mu_k$ (conditional on $\mu$,$\tau^2$) is given by the usual precision-based formulas, which we rewrite in a form more suggestive of shrinkage:
\begin{align}
\mu_k \mid m, \mu, \tau^2 &\sim \mathcal{N}\left(\frac{(\tau^2 - \sigma^2)^{-1} \mu + \sigma^{-2} m_k}{(\tau^2-\sigma^2)^{-1}+\sigma^{-2}}, \left[(\tau^2 - \sigma^2)^{-1} + \sigma^{-2}\right]^{-1}\right) \nonumber \\
&\sim \mathcal{N}\left(\mu + \left(1 - \frac{\sigma^2}{\tau^2}\right)(m_k - \mu), \left(1-\frac{\sigma^2}{\tau^2}\right)\sigma^2\right) \label{eq:posterior_basic}
\end{align}

To find unbiased point estimates of $\mu$ and $\tau^{-2}$ we use the marginal distribution of $m$:
\begin{equation}
m_k \mid \mu, \tau^2 \sim \mathcal{N}(\mu, \tau^2). \label{eq:marginal_y}
\end{equation}
Letting
\begin{align*}
\bar{m} &\coloneqq \frac{1}{K} \sum_{k=1}^K m_k &
s^2 &\coloneqq \sum_{k=1}^K (m_k - \bar{m})^2,
\end{align*}

we have
\begin{align*}
\E \bar{m} &= \mu &
\E\left(\frac{K-3}{s^2}\right) &= \frac{1}{\tau^2}
\end{align*}

Substituting into the posterior mean into \ref{eq:posterior_basic} yields the empirical Bayes estimator

\begin{align}
\hat{\mu}_k &= \bar{m} + \left( 1 - \frac{\sigma^2}{s^2 / (K - 3)}\right) (m_k - \bar{m}) \nonumber \\
&= \bar{y} + (1 - \xi) (m_k - \bar{m}) \label{eq:eb_basic}
\end{align}
the James-Stein estimator, where
$$
\xi \coloneqq \frac{\sigma^2}{s^2 / (K-3)}
$$
roughly the ratio of noise variance to sample variance of the observations (smaller by a factor of $(K-3) / (K - 1)$). 
When $\xi$ is close to one, most of the observed variance is explained by sampling noise, and we have good reason to shrink. 
When $\xi$ is close to zero, the samples are well spread out compared to sampling noise and we have little justification for shrinkage.

In practice, we restrict $\xi$ to be no greater than one, so that we don't ``shrink past the mean". The is called the positive-part James-Stein estimator.

From \ref{eq:posterior_basic}, the na\"ive estimator of posterior variance would be:
\begin{equation}
\mathbb{V}(\mu_k\mid m) = (1 - \xi)\sigma^2, \label{eq:naive_var_est}
\end{equation}

but this ignores the variability in $\bar{y}$ and $(K-3)/s^2$ as estimators of $\mu$ and $\tau^{-2}$.

\section{Fully Bayesian derivation}
\label{app:bayes_full}
Here we summarize a fully Bayesian calculation of the posterior variance $\mathbb{V}(\mu_k\mid m)$.

Instead of using point estimates of $\mu$ and $\tau^{-2}$, we assign improper uniform prior densities:
\begin{align*}
p(\tau^2) &\propto 1&
p(\mu\mid \tau^2) &\propto 1.
\end{align*}

We can show that the posterior distribution of $\mu, \tau^2$ is proper, given by
\begin{align}
\label{eq:posterior_mu} \mu \mid \tau^2, m &\sim \mathcal{N}\left(\bar{m}, \frac{\tau^2}{K}\right) \\ 
\tau^2 | m &\sim \textrm{Inverse-Gamma}\left(\frac{K-3}{2}, \frac{s^2}{2}\right). \nonumber
\end{align}

Actually, the posterior distribution of $\tau^2$ is a truncated inverse-gamma because $\tau^2 \in [\sigma^2, \infty)$ rather than $[0, \infty)$. We will ignore this point, using the gamma distribution to compute posterior moments of $\tau^{-2}$.

Taking posterior expectations over $\mu$ and then $\tau^2$ in the posterior mean for $\mu_k$, \ref{eq:posterior_basic} yields \ref{eq:eb_basic}, the James-Stein estimator, now as a fully Bayesian posterior expectation (ignoring the truncation of $\tau^2$). Furthermore, we can compute the posterior variance:

\begin{equation}
\mathbb{V}(\mu_k \mid m) \approx \underbrace{(1-\xi) \sigma^2}_{\substack{\text{uncertainty in $\mu_k$}}} + \overbrace{\frac{\xi \sigma^2}{K}}^{\mathclap{\substack{\text{uncertainty in $\mu$}}}}+  \underbrace{\frac{2 \xi^2 (m_k - \bar{m})^2}{K-3}}_{\substack{\text{uncertainty in $\tau^2$}}} \label{eq:var_complete}
\end{equation}

Appealing to the ``Bayesian CLT", we can use \ref{eq:posterior_basic} and \ref{eq:var_complete} to construct 95\% posterior intervals for each $\mu_k$. There are two sources of approximation error: ignoring the truncation of $\tau^2$ and ignoring non-normality of the posterior distribution. For full details on this derivation, see Appendix \ref{app:bayes_full}.

Formula (1.45) in \citet{efron2012} p. 11 is the corresponding version of \ref{eq:var_complete} with $\nu=0$ assumed. Efron says the derivation is from \citet{morris1983} where the variance expression appears as equation (4.1). The posterior mean and variance also appear in theorem 2.1 of \citet{vanderMerwe1988} along with cumbersome expressions for the third and fourth posterior moments.

Using \ref{eq:marginal_y}, the joint posterior of $\mu, \tau^2$ is given by
\begin{align*}
p(\mu, \tau^2 \mid m) &\propto p(\tau^2) p(\mu \mid \tau^2) p(m \mid \mu, \tau^2) \\
&\propto (\tau^2)^{-K/2} \exp\left\{-\frac{K(\bar{m} - \mu)^2}{2\tau^2} - \frac{s^2}{2\tau^2}\right\}.
\end{align*}

From this we see

$$
p(\mu\mid\tau^2, m) \propto \exp\left\{-\frac{(\mu - \bar{m})^2}{2(\tau^2 / K)}\right\},
$$
which shows that $\mu$ has a normal conditional posterior \ref{eq:posterior_mu}. 
Then the posterior of $\tau^2$ is given by
\begin{align*}
p(\tau^2\mid m) &\propto \frac{p(\mu, \tau^2)}{p(\mu\mid\tau^2,m)} \\
&\propto (\tau^2)^{-(K-1)/2}\exp\left\{-\frac{s^2}{2\tau^2}\right\}
\end{align*}
the density of an Inverse-Gamma$\left(\frac{K - 3}{2}, \frac{s^2}{2}\right)$ variable; hence
$$
\tau^{-2} \mid m \sim \textrm{Gamma}\left(\frac{K-3}{2}, \frac{s^2}{2}\right),
$$
ignoring truncation, and the approximate posterior moments of $\tau^{-2}$ are
\begin{align*}
\E(\tau^{-2} \mid m) &\sim \frac{K-3}{s^2} = \frac{\xi}{\sigma^2} \\
\mathbb{V}(\tau^{-2} \mid m) &\approx \frac{2(K-3)}{s^4} = \frac{2\xi^2}{(K-3)\sigma^4}.
\end{align*}

Returning to \ref{eq:posterior_basic}, the conditional posterior expectation of $\mu_k$ is

$$
\E(\mu_k \mid \mu, \tau^2, m) = \mu + \left(1 - \frac{\sigma^2}{\tau^2}\right) (m_k - \mu),
$$

and taking posterior expectations over $\mu$ and then $\tau^2$ yields \ref{eq:eb_basic}.For the variance, we use

$$
\mathbb{V}(\mu_k \mid m) = \E[\mathbb{V}(\mu_k\mid \mu, \tau^2, m)\mid m] + \mathbb{V}(\E[\mu_k\mid \mu, \tau^2, m]).
$$

The first term is

\begin{align}
\E[\mathbb{V}(\mu_k\mid\mu, \tau^2, m)\mid m] &= \E\left[ \left(1-\frac{\sigma^2}{\tau^2}\right)\sigma^2\mathrel{\Big|} m\right] \nonumber \\
&\approx (1 - \xi) \sigma^2, \label{eq:first_term}
\end{align}

the ``na\"ive'' variance, \ref{eq:naive_var_est}. The second term is

\begin{align}
\mathbb{V}(\E[\mu_k \mid \mu, \tau^2, m]\mid m) &= \mathbb{V}\left(\mu + \left(1-\frac{\sigma^2}{\tau^2}\right)(m_k - \mu)\mathrel{\Big|} m\right) \nonumber \\
&= \sigma^4 \mathbb{V}\left(\frac{m_k-\mu}{\tau^2} \mathrel{\Big|} m \right) \nonumber \\
&= \sigma^4\left[\E\left[\frac{1}{K\tau^2} \mathrel{\Big|} m\right] + \mathbb{V}\left(\frac{m_k - \bar{m}}{\tau^2} \mathrel{\Big|} \right) \right] \tag{using \ref{eq:posterior_mu}} \nonumber \\
&\approx \frac{\sigma^2\xi}{K} + \frac{2\xi^2(m_k - \bar{m})^2}{K-3}. \label{eq:second_term}
\end{align}

Summing \ref{eq:first_term} and \ref{eq:second_term} yields \ref{eq:var_complete}. 

\subsection{A mixture view of James-Stein estimation}

It is interesting that the Stein estimator looks a lot like a mixture between two posteriors: one looking at the component alone,
$$
\mu_k \mid m \sim \mathcal{N}(m_k, \sigma^2),
$$

and one treating the entire ensemble as coming from a single mean,
$$
\mu_k \mid m \sim \mathcal{N}\left(\bar{m}, \frac{\sigma^2}{K}\right)
$$

with mixture weight $\xi$ on the ensemble and $(1-\xi)$ on the component. The expectation of such a mixture is exactly the same as the James-Stein estimator, \ref{eq:eb_basic}, and the variance of such a mixture is
$$
(1-\xi)\sigma^2 + \frac{\sigma^2\xi}{K} + \xi (1-\xi)(m_k - \bar{m})^2,
$$
which is similar to \ref{eq:var_complete}, but not quite the same.

\subsection{An alternative prior}

We might alternatively set $p(\tau^2) \propto \frac{1}{\tau^2}$,
an ``uninformative'' choice which can be motivated by a pivot argument (see \citet[p. 54]{gelman2014}, though here $\tau$ is not a pure scale parameter). This prior simply replaces the $K-3$ by a $K-1$: letting
$$
\tilde{\xi} \coloneqq \frac{\sigma^2}{s^2 / (K-1)},
$$
we have
\begin{align*}
\E(\mu_k\mid m) &\approx \bar{y} + (1-\tilde{\xi})(m_k - \bar{m}) \\
\mathbb{V}(\mu_k \mid m) &\approx (1-\tilde{\xi})\sigma^2 + \frac{\sigma^2\tilde{\xi}}{K} + \frac{2\tilde{\xi}^2(m_k - \bar{m})^2}{K-1}
\end{align*}

\end{document}